\documentclass[PRL,noshowpacs,amsmath,amssymb,superscriptaddress,twocolumn]{revtex4}
\usepackage[T1]{fontenc}
\usepackage[latin1]{inputenc}
\usepackage{amsmath}
\usepackage{amssymb}
\usepackage{graphicx}
\usepackage{epsfig}
\usepackage{bm}
\usepackage{dsfont}
\usepackage{color}

\newcommand{\bea}{\begin{eqnarray}}
\newcommand{\eea}{\end{eqnarray}}
\newcommand{\be}{\begin{equation}}
\newcommand{\ee}{\end{equation}}
\newcommand{\bc}{\begin{center}}
\newcommand{\ec}{\end{center}}

\newcommand{\bsig}{\boldsymbol\sigma}                                                                 
                      
\newcommand{\bwt}{\begin{widetext}}  
\newcommand{\ewt}{\end{widetext}}
 
\newcommand{\Eq}[1]{Eq.~(\ref{#1})}

\newcommand{\phdagger}{{\phantom{\dagger}}}

\begin{document}

\title{Kondo Destruction in RKKY-Coupled Kondo Lattice
and Multi-Impurity Systems}

\author{Ammar Nejati}
\affiliation{Physikalisches Institut and Bethe Center for Theoretical Physics,
Universit\"at Bonn, Nussallee 12, D-53115 Bonn, Germany}

\author{Katinka Ballmann}
\affiliation{Physikalisches Institut and Bethe Center for Theoretical Physics,
  Universit\"at Bonn, Nussallee 12, D-53115 Bonn, Germany}

\author{Johann Kroha}
\email[Email: ]{kroha@physik.uni-bonn.de}

\affiliation{Physikalisches Institut and Bethe Center for Theoretical Physics,
  Universit\"at Bonn, Nussallee 12, D-53115 Bonn, 
Germany} 

\affiliation{Center for Correlated Matter, Zhejiang University, 
Hangzhou, Zhejiang 310058, China} 

\received{December 8, 2016; published March 8, 2017}

\begin{abstract}
In a Kondo lattice, the spin exchange coupling between a local spin 
and the conduction electrons acquires nonlocal contributions due to 
conduction electron scattering from surrounding local spins and the
subsequent RKKY interaction. It leads to a hitherto unrecognized interference 
of Kondo screening and the RKKY interaction beyond the Doniach scenario.
We develop a renormalization group theory for the RKKY-modified 
Kondo vertex. The Kondo temperature, $T_K(y)$, is suppressed in a 
universal way, controlled by the dimensionless 
RKKY coupling parameter $y$. Complete spin screening ceases to exist 
beyond a critical RKKY strength $y_c$ even in the absence of magnetic
ordering. At this breakdown point, $T_K(y)$ remains nonzero and
is not defined for larger RKKY couplings, $y>y_c$. The results are in
quantitative agreement with STM spectroscopy experiments on 
tunable two-impurity Kondo systems. The possible implications for 
quantum critical scenarios in heavy-fermion systems are discussed.

\end{abstract}


\maketitle

The concept of fermionic quasiparticles existing even in strongly
interacting many-body systems is fundamental for 
a wealth of phenomena summarized under the term 
Fermi liquid physics. In heavy-fermion systems \cite{Loehneysen07}, 
quasiparticles with a large effective mass are formed by the Kondo effect
\cite{Hewson93}.
The conditions under which these heavy quasiparticles disintegrate near 
a quantum phase transition (QPT) have been an 
important, intensively debated and still open issue for many years
\cite{Loehneysen07}.

The heavy Fermi liquid, like any other Fermi liquid, may undergo a 
spin density-wave (SDW) instability, leading to critical fluctuations of the 
magnetic order parameter but leaving the heavy quasiparticles intact.
This scenario is well described by the pioneering works of Hertz, 
Moriya and Millis \cite{Hertz76,Moriya85,Millis93}. However, early on Doniach 
pointed out \cite{Doniach77} that the Kondo spin screening of the 
local moments should eventually cease and give way to magnetic order, 
when the RKKY coupling energy between the local moments 
\cite{Ruderman54,Kasuya56,Yosida57} becomes larger than the characteristic
energy scale for Kondo singlet formation, the Kondo temperature $T_K$. 
It is generally believed that the Kondo destruction is driven by the 
critical fluctuations near a QPT. Several mechanisms have been proposed, 
invoking different types of fluctuations, including critical fluctuations 
of the local magnetization coupling to the 
fermionic quasiparticles (local quantum criticality) 
\cite{Si01,Coleman01} and Fermi surface fluctuations self-consistently 
generated by the Kondo destruction \cite{Senthil04}. 
Most recently, a scenario of critical quasiparticles with diverging effective 
mass and a singular interaction, induced by critical antiferromagnetic 
fluctuations, has been put forward \cite{Woelfle11,Woelfle14, Woelfle16}. 
Intriguing in its generality, it does, however, not invoke Kondo physics.

Here, we show that the heavy-electron quasiparticles can be destroyed by
the RKKY interaction even without critical fluctuations. This occurs
because of a hitherto unrecognized feedback effect: in a 
Kondo lattice or multi-impurity system, the RKKY interaction,
parametrized by a dimensionless coupling $y$, reduces the 
Kondo screening energy scale $T_K(y)$. This reduction  
implies an increase of the local spin susceptibility at low temperatures $T$, 
$\chi _f(T=0)\sim 1/T_K(y)$, which in turn increases the effective
RKKY coupling. We derive this effect and analyze it 
by a renormalization group (RG) treatment. In particular, we calculate the 
temperature scale for Kondo singlet formation in a Kondo lattice, $T_K(y)$.
It is suppressed with increasing $y$ in a universal way. 
Beyond a critical RKKY coupling $y_c$, complete Kondo singlet formation 
ceases to exist. However, at this breakdown point $T_K(y_c)$ remains finite, 
and the suppression with respect to the single-impurity Kondo scale takes 
a universal value, $T_K(y_c)/T_K(0)=1/{e}$, where ${e}=2.718...$ 
is Euler's constant. 
These findings are consistent with conformal field theory results 
\cite{Affleck95,Sela09} and in quantitative agreement with STM spectroscopy 
experiments on tunable, RKKY-coupled two-impurity Kondo systems 
\cite{Bork11,Kroeger11}. 

The present results directly apply to cases where long-range order 
does not play a role, that is, two-impurity Kondo systems 
\cite{Bork11,Kroeger11,Prueser14}, 
compounds where the magnetic ordering does not occur at the Kondo breakdown 
point \cite{Friedemann09}, and temperatures sufficiently above the 
N\'eel temperature \cite{Klein08}. 
They will set the stage for a complete theory 
of heavy-fermion quantum criticality by including critical order-parameter
fluctuations either of the incompletely screened magnetic moments or 
of an impending SDW instability.

\begin{figure}[t]
\includegraphics[width=0.85\linewidth]{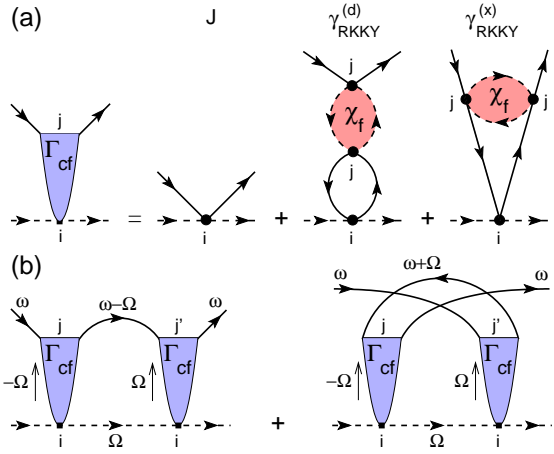}
\caption{\label{fig:vertex} 
(a) $f$-spin--$c$-electron vertex $\hat\Gamma_{cf}$, composed of the on-site 
vertex $J$ at site $i$ and the RKKY-induced contributions from 
surrounding sites $j\neq i$ to leading order in the RKKY coupling: 
$\gamma_{{\rm RKKY}}^{(d)}$ (direct term) and $\gamma_{{\rm RKKY}}^{(x)}$ (exchange term).   
(b) One-loop diagrams for the perturbative RG. 
Solid lines: electron Green's functions $G_c$.
Dashed lines: pseudofermion propagators $G_f$ of the local $f$ spins. 
The red bubbles represent the full $f$-spin susceptibility 
at sites $j$.  }
\end{figure}

{\it The model.} -- We consider the Kondo lattice model  
\be
H = \sum_{{\bf k},\sigma} 
\varepsilon_{\bf k} \; c_{{\bf k} \sigma}^\dagger c_{{\bf k} \sigma}^\phdagger +
J_0 \sum_{i} {\bf S}_i \cdot {\bf s}_i 
\label{Hamiltonian}
\ee
where $c_{{\bf k} \sigma}^\phdagger$, $c_{{\bf k}\sigma }^\dagger$ denote the conduction 
($c$) electron operators with dispersion $\varepsilon_{\bf k}$. ${\bf S}_i$
are the local spin operators at the lattice sites ${\bf x}_i$, 
exchange coupled to the conduction electron spins
${\bf s}_i = \sum_{\sigma,\sigma'}  c_{i \sigma}^\dagger \bsig_{\sigma\sigma '}
c_{i\sigma '}^\phdagger$ via an on-site, antiferromagnetic coupling $J_0>0$.
The local spins will henceforth be termed $f$ spins, as they are
typically realized in heavy fermion systems by the rare-earth $4f$ 
electrons. We will use the pseudofermion  
representation of the $f$ spins, ${\bf S}_i = 
1/2\ \sum_{\tau,\tau'}f_{i \tau}^\dagger \bsig_{\tau\tau '} f_{i \tau '}^\phdagger$,
with $\bsig$ the vector of Pauli matrices and  
$f_{i \sigma}^\phdagger$, $f_{i\sigma '}^\dagger$ fermionic operators obeying the 
constraint $\hat Q=\sum_{\sigma} f_{i \sigma}^\dagger f_{i\sigma}^\phdagger = 1$.
It is crucial that the coupling between different $f$ spins is not a direct 
exchange interaction, but mediated by the conduction band 
\cite{Ruderman54,Kasuya56,Yosida57} and generated in second 
order by the same spin coupling $J_0$ that also creates the Kondo 
effect. The essential difference can be seen from the example of a 
two-impurity Kondo system, ${\bf S}_1$, ${\bf S}_2$: 
with a direct impurity-impurity coupling $K\, {\bf S}_1 \cdot {\bf S}_2 $, 
and for a specific particle-hole symmetry \cite{Affleck95},
this model can exhibit a dimer singlet phase where the 
dimer is decoupled from the conduction electrons (scattering phase shift 
at the Fermi energy $\phi_{{\rm dimer}} =0$). As a function of $K$, 
this dimer singlet phase is then separated from the Kondo singlet 
phase (scattering phase shift $\phi_{{\rm Kondo}}=\pi/2$) by a quantum critical 
point (QCP) \cite{Jones88,Affleck95}, see also Ref.~\cite{Sela09}.   
By contrast, when the interimpurity coupling is controlled by the RKKY 
interaction only, i.e. generated by $J_0$, a decoupled dimer singlet
and, hence, a second-order QCP is not possible. Instead, we find below 
that the Kondo singlet formation at $T=0$ breaks down at a critical 
strength of the RKKY coupling, however without a diverging local 
impurity susceptibility, that is, with a discontinuous jump of $T_K(y)$.
The profound implications of this behavior will be discussed below.

{\it RKKY-coupled c-f vertex and renormalization group}.-- 
We develop an analytical renormalization group for 
RKKY-coupled Kondo multi-impurity and lattice systems, taking 
the proper renormalizations of all appearing vertices into account. 
The RKKY vertex $\hat\Gamma_{ff}$ coupling two $f$ spins has no logarithmic
RG flow, since the recoil (momentum integration) of the itinerant conduction 
electrons prevents an infrared singularity of the RKKY interaction. 
$\hat\Gamma_{ff}$ thus remains in the weak coupling regime. 
The formation of the strong-coupling Kondo singlet, which is the origin 
of the heavy-Fermion state, is signalled by a RG divergence of the  
spin-scattering vertex operator $\hat\Gamma_{cf}$ between $c$ electrons 
and an $f$ spin. In the case of multiple Kondo sites, this vertex 
acquires nonlocal contributions in addition to the local coupling $J$ at 
a site $i$, because a $c$ electron can scatter from a distant Kondo site 
$j\neq i$, and the spin flip at that site is transferred to the $f$ spin 
at site $i$ via the RKKY interaction. In this way, $\hat\Gamma_{ff}$ will 
influence the RG flow of $\hat\Gamma_{cf}$, even though it is not renormalized 
itself. The corresponding diagrams are shown in Fig.~\ref{fig:vertex} (a). 
As seen from the figure, such a nonlocal scattering process necessarily 
involves the exact, local dynamical $f$-spin susceptibility $\chi_f(i\Omega)$ 
on site $j$. The resulting $c-f$ vertex $\hat\Gamma_{cf}$ has the 
structure of a nonlocal Heisenberg coupling in spin space. 
The exchange diagram, $\gamma_{{\rm RKKY}}^{(x)}$ in Fig.~\ref{fig:vertex} (a), 
contributes only a subleading logarithmic term as compared to
$\gamma_{{\rm RKKY}}^{(d)}$ \cite{supplement}. In particular, it does not alter the 
universal $T_K(y)$ suppression derived below and can, therefore, be
neglected. To leading (linear) order in the RKKY coupling, 
$\hat\Gamma_{cf}$ thus reads (in Matsubara representation), 
\begin{eqnarray}
\hat\Gamma_{cf} &=& 
\left[ J  \delta_{ij} + \gamma^{(d)}_{RKKY}({\bf r}_{ij},i\Omega)\right] 
{\bf S}_i \cdot {\bf s}_j 
\label{eq:Gammacf}\\ &&\hspace*{-1.2cm}
= \left[ J\delta_{ij}+ 2J J_0^2\ (1-\delta_{ij})\ \chi_c({\bf r}_{ij}, i\Omega)\, 
\tilde\chi_f(i\Omega) \right] 
{\bf S}_i \cdot {\bf s}_j \, ,
\nonumber
\end{eqnarray}
where ${\bf r}_{ij} = {\bf x}_i - {\bf x}_j$ is the distance vector 
between the sites $i$ and $j$, and $\Omega$ is the energy transferred in the 
scattering process. $\chi_c({\bf r}_{ij}, i\Omega)$ is the $c$ electron 
density correlation function between sites $i$ and $j$ [bubble of solid lines 
in Fig.~\ref{fig:vertex} (a)] and 
$\tilde\chi_f(i\Omega):=\chi_f(i\Omega)/(g_L\mu_B)^2$ with $g_L$ the 
Land\'e factor and $\mu_B$ the Bohr magneton. 
Note that \Eq{eq:Gammacf} contains the running coupling 
$J$ at site $i$, which will be renormalized under the RG, 
while at the site $j$, where the $c$ electron scatters, the bare 
coupling $J_0$ appears, since all vertex renormalizations on that 
site are already included in the exact susceptibility $\chi_f$. 
Higher order terms, as for instance generated by the RG
[see below, Fig.~\ref{fig:vertex} (b)], lead to nonlocality of the incoming 
and outgoing coordinates of the scattering $c$ electrons, 
${\bf x}_{j}$, ${\bf x}_{j'}$, but the $f$-spin coordinate ${\bf x}_i$ 
remains strictly local, since the pseudofermion propagator 
$G_f(i\nu)=1/i\nu$ is local \cite{Kroha98}. 
For this reason, speaking of Kondo singlet formation on 
a single Kondo site is well defined even in a Kondo lattice, and so is 
the local susceptibility $\chi_f$ of a single $f$ spin.
The corresponding Kondo scale $T_K$ on a site $j$ is observable, 
e.g., as the Kondo resonance width measured by STM spectroscopy on one Kondo 
ion of the Kondo lattice.
The temperature dependence of the single-site $f$-spin susceptibility 
is known from the Bethe ansatz solution \cite{Andrei83} in terms of the 
Kondo scale $T_K$. It has a 
$T=0$ value $\chi_f(0)\propto 1/T_K$ and crosses over to the $1/T$ 
behavior of a free spin for $T>T_K$. These features can be 
modeled in the retarded or advanced, local, dynamical  
$f$-spin susceptibility $\chi_f(\Omega\pm i0)$ as
\begin{eqnarray}
\chi_f(\Omega \pm i0) &=& \frac{(g_L\mu_B)^2W}{\pi T_K\,\sqrt{1+(\Omega/T_K)^2}}\, 
\left( 1 \pm \frac{2i}{\pi} {\rm arsinh} \frac{\Omega}{T_K} \right)
\nonumber\\
\label{eq:chif}
\end{eqnarray} 
where $W$ is the Wilson ratio, and the imaginary part is implied by the 
Kramers-Kronig relation. 

We now derive the one-loop RG equation for the $c-f$ vertex $\hat\Gamma_{cf}$,
including RKKY-induced, nonlocal contributions.
The one-loop spin vertex function is shown diagrammatically 
in Fig.~\ref{fig:vertex} (b). Using Eq.~(\ref{eq:Gammacf}), the sum of these
two diagrams is up to linear order in the RKKY coupling 
\begin{eqnarray}
Y({\bf r}_{ij}, i\omega ) &=&   \label{eq:Yx} \\  
&&\hspace*{-2.1cm} -
J \, T \sum_{i\Omega}   
\left[  J\delta_{ij} +
\gamma^{(d)}_{RKKY}({\bf{r}}_{ij},i\Omega)+\gamma^{(d)}_{RKKY}({\bf{r}}_{ij},-i\Omega)
\right] \nonumber \\
&&\hspace*{-0.9cm}
\times 
\left[ 
G_c({\bf r}_{ij}, i\omega -i\Omega )-G_c({\bf r}_{ij}, i\omega + i\Omega)
\right]\, G_f(i\Omega). 
\nonumber
\end{eqnarray}
Here, $\omega$ is the energy of the incoming conduction electrons,    
$G_c({\bf r}_{ij}, i\omega + i\Omega)$ is the single-particle 
$c$-electron propagator from the incoming to the outgoing site. 
For example, for an isotropic system, 
$G_c({\bf r},\omega \pm i0) = - \pi N(\omega)\,
{{\rm e}^{\pm i k(\varepsilon_F+\omega)r}}/
{k(\varepsilon_F+\omega)r}$, with the bare density of states
$N(\omega)$, and $k(\varepsilon_F+\omega)$ the modulus of the 
momentum corresponding to the energy $\omega$.

\begin{figure}[t] 
\includegraphics[width=0.95\linewidth]{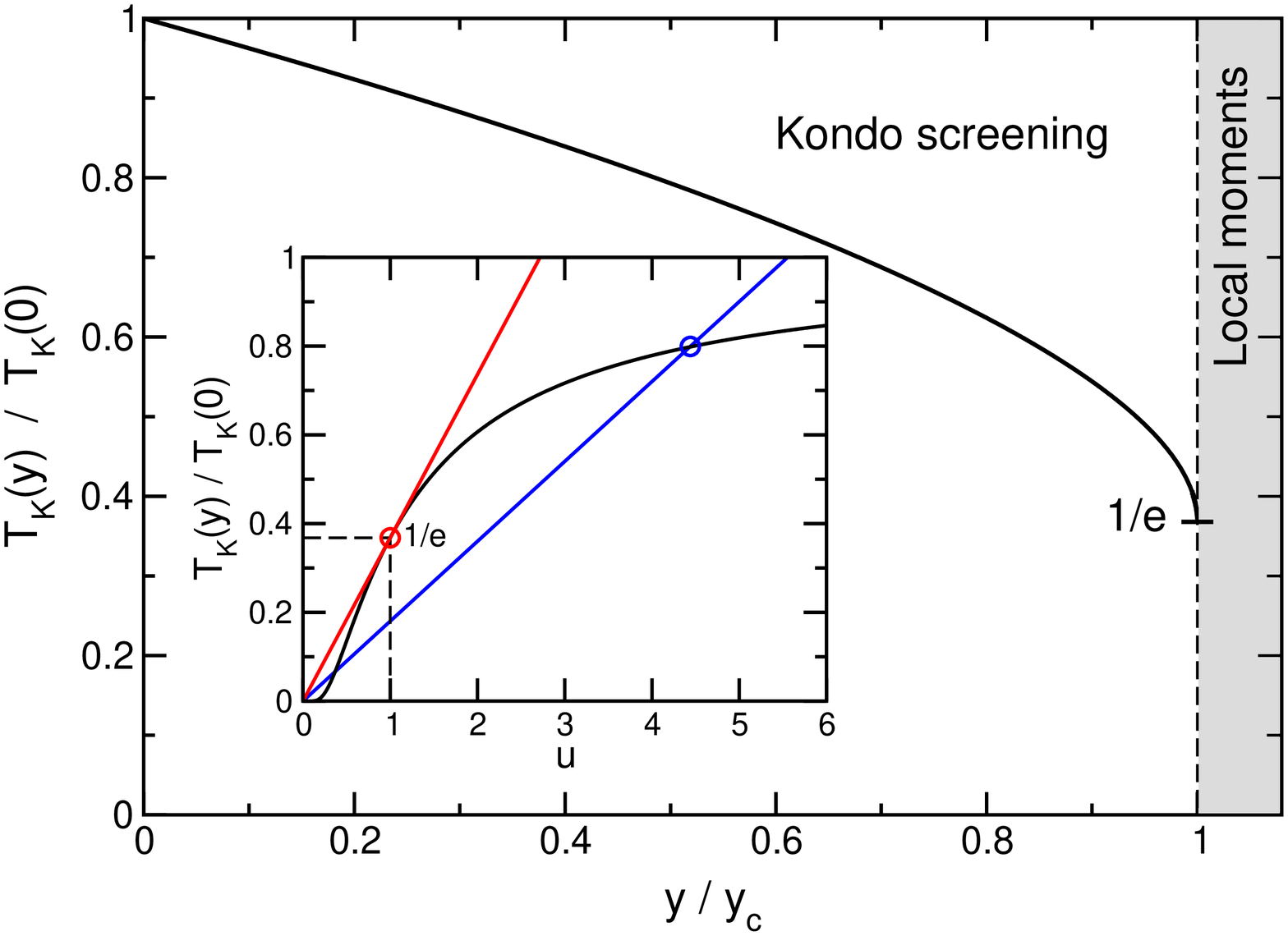}
\caption{\label{fig:TK_y_universal}
Universal dependence of $T_K(y)/T_K(0)$ on the normalized 
RKKY parameter $y/y_c$, solution of \Eq{eq:TK_y_universal}. 
The inset visualizes the solution of \Eq{eq:TK_y_universal} graphically.
Black, solid curve: right-hand side of \Eq{eq:TK_y_universal}. Blue line: 
left-hand side for $y<y_c$. Red line: left-hand side for $y=y_c$ (where 
the red line and black curve touch). It proves that there is a
critical coupling $y_c$ beyond which \Eq{eq:TK_y_universal} has no solution,
and $T_K(y_c)/T_K(0)=1/e$.
\vspace*{-0.0cm}}   
\end{figure}

For the low-energy physics, the vertex renormalization for $c$ electrons 
at the Fermi surface is required.   
This means setting the energy $i\omega\to\omega =0+i0$  and 
Fourier transforming the total vertex 
$Y({\bf r}_{ij}, i\omega)$ with respect to the incoming and outgoing 
$c$ electron coordinates, ${\bf x}_j$, ${\bf x}_i$, and taking 
its Fourier component for momenta at the Fermi surface ${\bf k}_F$,
see Ref.~\cite{supplement}. Note that at the Fermi energy  
$Y({\bf k}_F, 0)$ is real, even though the RKKY-induced, 
dynamical vertex $\gamma^{(d)}_{{\rm RKKY}}(\pm i\Omega)$ appearing in 
\Eq{eq:Yx} is complex valued \cite{supplement}. This ensures  
the total vertex operator of the renormalized Hamiltonian is
Hermitian. By analytic continuation, the Matsubara summation
in \Eq{eq:Yx} becomes an integration over 
the intermediate $c$ electron energy from the lower and upper 
band cutoff $D$ to the Fermi energy ($\Omega=0$). The 
coupling constant renormalization is then obtained in the standard way 
by requiring that $Y({\bf k}_F, 0)$ be invariant under an infinitesimal 
reduction of the running band cutoff $D$.
Note that the band cutoff appears in both, the
intermediate electron propagator $G_c$ and in $\chi_c$.
However, differentiation of the latter does not contribute to 
the logarithmic RG flow. This leads to the one-loop RG equation
\cite{supplement} 
\be
\frac{{d}g}{{d}\ln  D} = -2 g^2\ 
\left( 1 - y\, g_0^2 \ \frac{D_0}{T_K}\ \frac{1}{\sqrt{1+(D/T_K)^2}} 
\right) \ ,
\label{eq:RGequation}
\ee
where we have introduced the dimensionless couplings $g=N(0)J$,
$g_0=N(0)J_0$, and $D_0$ is the bare band cutoff. 
The first term on the right-hand side of \Eq{eq:RGequation} is the on-site 
contribution to the differential coupling renormalization 
(the $\beta$ function), while the second term 
represents the RKKY contribution. It is seen that $\chi_f$, as in \Eq{eq:chif},
induces a soft cutoff on the scale $T_K$ and the characteristic
$1/T_K$ dependence to the RG flow of this contribution, where 
$T_K$ is the Kondo scale on the {\it surrounding} Kondo sites.      
The dimensionless coefficient 
\be
y=-\frac{8W}{\pi ^2} {\rm Im} \sum_{j\neq i}
\frac{{\rm e}^{-i{\bf k}_F{\bf r}_{ij}}} {N(0)^2}
 G_c^R({\bf r}_{ij}, \Omega=0 ) \,
\chi_c({\bf r}_{ij}, \Omega=0 ) 
\label{eq:RKKYparameter1}
\ee
arises from the Fourier transform $Y({\bf k}_F, 0)$ and  
parametrizes the RKKY coupling strength. The summation 
in \Eq{eq:RKKYparameter1} runs over all positions $j\neq i$ 
of Kondo sites in the system. It is important to note that  
$y$ is generically positive \cite{supplement}, 
even though the RKKY correlations 
$\chi_c({\bf r}_{ij},0)$ may be ferro- or antiferromagnetic. 
For instance, for an isotropic and dense system with lattice constant 
$a$ ($k_Fa \ll 1$), 
the summation in Eq.~(\ref{eq:RKKYparameter1}) can be approximated 
by an integral, and with the substitution $x=2k_F|{\bf r}_{ij}|$, $y$ 
can be expressed as 
\be
y \approx  \frac{2W}{(k_Fa)^3}
\int_{k_Fa}^{\infty} dx\ (1-\cos x)\ \frac{x \cos x - \sin x}{x^4} > 0 \ .
\label{eq:RKKYparameter2}
\ee 
As a consequence, the RKKY correlations reduce the $g-$renormalization 
in \Eq{eq:RGequation}, irrespective of the sign of $\chi_c({\bf r}_{ij},0)$,
as one would physically expect.

{\it Universal suppression of the Kondo scale.} --
The RG (\ref{eq:RGequation}) can  be integrated 
analytically \cite{supplement}. 
The Kondo scale for singlet formation on site $i$ 
is defined as the running cutoff value where the $c-f$ coupling $g$ diverges.
By equivalence of all Kondo sites, this is equal to the Kondo scale $T_K$ 
on the surrounding sites $j\neq i$, which appears as a parameter 
in the $\beta$ function on the right-hand side of \Eq{eq:RGequation}.
This implies an implicit equation for the Kondo scale $T_K=T_K(y)$ 
in a Kondo lattice, and that it depends  
on the RKKY parameter $y$

\vspace*{-0.1cm}
\be
\frac{T_K(y)}{T_K(0)} = 
\exp \left( -y\, \alpha\, g_0^2\, \frac{D_0}{T_K(y)}  
\right) \ .
\label{eq:TK_y}
\ee
\vspace*{-0.1cm}

\noindent
Here, $T_K(0)=D_0\exp (-1/2g_0)$ is the single-ion Kondo scale without 
RKKY coupling, and \hbox{$\alpha =  \ln (\sqrt{2}+1)$}. By the rescaling,
$u=T_K(y)/(y\alpha g_0^2D_0)$, $y_c=T_K(0)/(\alpha {e} g_0^2D_0)$, 
\Eq{eq:TK_y} takes the universal form ($e$ is Euler's constant),

\vspace*{-0.2cm}
\be
\frac{y}{{e}y_c}\, u={e}^{-1/u} \ .
\label{eq:TK_y_universal}
\ee
\vspace*{-0.2cm}

\noindent
Its solution can be expressed in terms of the Lambert $W$ function 
\cite{Lambert} as $u(y)=-1/W(-y/{e}y_c)$. 
The inset of Fig.~\ref{fig:TK_y_universal}
visualizes solving \Eq{eq:TK_y_universal} graphically. It shows 
that \Eq{eq:TK_y_universal} has solutions only for $y\leq y_c$. This means that 
$y_c$ marks a Kondo breakdown point beyond which the RG does not scale 
to strong coupling; i.e., a Kondo singlet is not formed for $y>y_c$ 
even at the lowest energies. 
Using the above definitions, the RKKY-induced suppression of the Kondo 
lattice temperature reads 
$T_K(y)/T_K(0)= u(y)y/({e}y_c)=-y/[{e}y_cW(-y/{e}y_c)]$. It is
shown in Fig.~\ref{fig:TK_y_universal}. In particular, at the breakdown 
point it vanishes {\it discontinuously} and takes the finite, 
universal value (see inset of Fig.~\ref{fig:TK_y_universal}),
\bea
\frac{T_{K}(y_c)}{T_K(0)}=\frac{1}{{e}}\approx 0.368\ . 
\label{eq:TK_yc}
\eea
We emphasize that the RKKY parameter $y$ depends on details of  
the conduction band structure, including band renormalizations 
caused by the Kondo effect (coupling to the heavy-fermion band).
It also depends on the spatial arrangement of Kondo sites.
Subleading contributions to $\Gamma_{cf}$ may modify the form of the cutoff 
function in the RG~(\ref{eq:RGequation}) and thus the nonuniversal 
parameter $\alpha$. However, all this does not affect the universal 
dependence of $T_K(y)$ on $y$ given by \Eq{eq:TK_y_universal}.

The critical RKKY parameter, as defined before \Eq{eq:TK_y_universal}, 
can be expressed solely in terms of the single-ion Kondo scale

\vspace*{-0.25cm}
\be
y_{c} = \frac{4} {\alpha {e}}\tau_K ({\rm ln}\tau_K)^2 
\label{eq:yc}
\ee
\vspace*{-0.2cm}

\noindent 
with $\tau_K=T_K(0)/D_0$. 
Note that [via $T_K(0)=D_0 \exp (-1/2g_0)$ and $N(0)=1/(2D_0)$] this is 
equivalent to Doniach's breakdown criterion \cite{Doniach77}, 
$N(0)y_cJ_0^2 = T_K(0)$ up to a factor of $O(1)$. However, the present 
theory goes beyond the Doniach scenario in that it predicts the 
behavior of $T_K(y)$. 
 
\begin{figure}[t]
\includegraphics[width=0.95\linewidth]{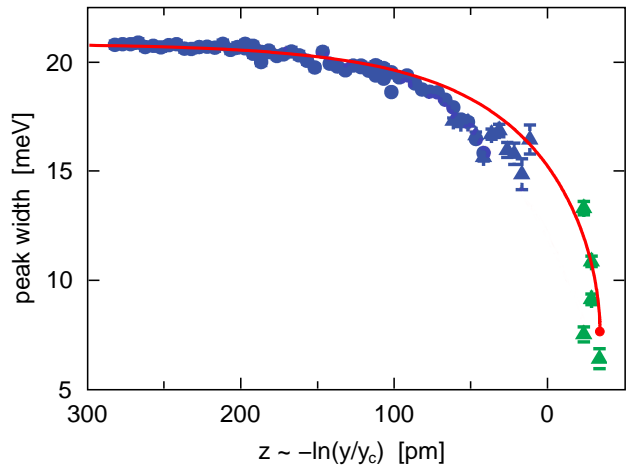}
\vspace*{-0.35cm}
\caption{\label{fig:TK_y_exp_th} Comparison of the theory
(\ref{eq:TK_y_universal}) (red curve) with STM spectroscopy experiments 
on a tunable two-impurity Kondo system \cite{Bork11} (data points).  
The data points represent the Kondo scale $T_K$ as extracted from the STM
spectra by fitting a split Fano line shape of width $T_K$ to the 
experimental spectra. Blue points: STM tunneling regime. 
Green points: contact regime. See Ref.~\cite{Bork11} 
for experimental details.} 
\end{figure}

{\it Comparison with experiments.} --
The present theory applies directly to two-impurity 
Kondo systems and can be compared to corresponding STM experiments 
\cite{Bork11,Kroeger11}. In Ref.~\cite{Bork11}, 
the Kondo scale has been extracted as the 
line width of the (hybridization-split) Kondo-Fano resonance. In this 
experimental setup, the RKKY parameter $y$ is proportional to the 
overlap of tip and surface $c$ electron wave functions and, thus,
depends exponentially on the tip-surface separation $z$, 
$y=y_c\,\exp[-(z-z_0)/\xi]$. Identifying the experimentally observed
breakdown point $z=z_0$ with the Kondo breakdown point, 
the only adjustable parameters are a scale factor $\xi$ of the $z$ coordinate
and $T_K(0)$, which is the resonance width at large separation, 
$z=300$~pm. 
The agreement between theory and experiment is striking, as shown in
Fig.~\ref{fig:TK_y_exp_th}. In particular, at the breakdown point
$T_K(y_c)/T_K(0)$ coincides accurately with the prediction 
(\ref{eq:TK_yc}) without any adjustable parameter. 
In the STM experiment of Ref.~\cite{Kroeger11},
the strongest observed suppression ratio is $T_K(y)/T_K(0) = 
46\,{\rm K}/110\,{\rm K}\approx 0.42$, again in excellent agreement with the 
strongest theoretical suppression of $1/e$, considering that 
in Ref.~\cite{Kroeger11} the RKKY coupling $y$ cannot be varied continuously.
The detailed analysis of that experiment will be published 
\hbox{elsewhere \cite{Nejati16}.}

{\it Discussion and conclusion.} --
We have derived a perturbative renormalization group theory 
for the interference of Kondo singlet formation and 
RKKY interaction in Kondo lattice and multi-impurity systems,
assuming that magnetic ordering is suppressed, e.g., by frustration. 
The equivalence of the $c-f$ vertices on all Kondo sites is reminiscent 
of a dynamical mean-field theory treatment; however, it goes beyond 
the latter in taking the nonlocal RKKY contributions into account.   
Equations~(\ref{eq:TK_y}) or (\ref{eq:TK_y_universal}) represent a mathematical 
definition of the energy scale for Kondo singlet formation in a Kondo lattice,
i.e., of the Kondo lattice temperature $T_K(y)$.
The theory predicts a universal suppression of $T_K(y)$ and a breakdown of 
complete Kondo screening at a critical RKKY parameter $y=y_c$. 
At the breakdown point, the Kondo scale takes a {\it finite}, universal value,
$T_K(y_c)/T_K(0)=1/{e}\approx 0.368$, and vanishes 
{\it discontinuously} for $y>y_c$. 
In the Anderson lattice, by contrast to the Kondo lattice, the locality 
of the $f$ spin no longer strictly holds, but our approach should 
still be valid in this case. The parameter-free, quantitative agreement 
of this behavior with different spectroscopic experiments 
\cite{Bork11,Kroeger11} strongly supports that the present theory 
captures the essential physics of the Kondo-RKKY interplay. 

The results may have profound relevance for heavy-fermion magnetic QPTs. 
In an unfrustrated lattice, the partially screened 
local moments existing for $y>y_c$ must undergo a second-order 
magnetic ordering transition at sufficiently low temperature. 
This will also imply a power law divergence of the $c$ electron correlation 
$\chi_c$ in \Eq{eq:Gammacf}. We have checked the effect of such a  
magnetic instability, induced either by the ordering of remanent 
local moments or by a $c-$electron SDW instability: 
the breakdown ratio $T_K(y_c)/T_K(0)$ will be altered, but must 
remain nonzero. The reason is that the inflection point of the exponential 
function on the right-hand side of \Eq{eq:TK_y_universal} 
(see Fig.~\ref{fig:TK_y_universal}) is not changed by such 
a divergence and, therefore, the solution ceases to exist at a finite 
value of $T_K(y_c)$.
This points to an important conjecture about a possible, new quantum 
critical scenario with Kondo destruction: the Kondo spectral weight may
vanish continuously at the QCP, while the Kondo scale $T_K(y)$ 
(resonance width) remains finite, both as observed experimentally in 
Ref.~\cite{Bork11}.
Such a scenario may reconcile apparently contradictory experimental results
in that it may fulfill dynamical scaling, even  
though $T_K(y_c)$ is finite at the QCP. 
The present theory sets the stage for constructing a complete
theory of magnetic ordering and RKKY-induced Kondo destruction. 

We gratefully acknowledge useful discussions with Manfred Fiebig,
Stefan Kirchner, J\"org Kr\"oger, Nicolas N\'eel, 
Hilbert von L\"ohneysen, Qimiao Si, Matthias Vojta, Peter Wahl and
Christoph Wetli. We especially thank Peter 
Wahl for allowing us to use the experimental STM data of Ref. \cite{Bork11}. 
This work was supported in part (J.K., A.N.)
by the Deutsche Forschungsgemeinschaft (DFG) through Grant No. SFB-TR185.


\begin{widetext}

\vfill
\eject

\begin{centering}
{\large{\bf Supplemental Material for\\[0.2cm]
``Kondo destruction in RKKY-coupled Kondo lattice 
and multi-impurity systems''}}\\[0.3cm]

Ammar Nejati,$^1$ Katinka Ballmann,$^1$ Johann Kroha$^{1,2}$ \\
{\it $^1$ Physikalisches Institut and Bethe Center for Theoretical Physics,\\ 
Universit\"at Bonn, Nussallee 12, D-53115 Bonn, Germany}\\
{\it $^2$ Center for Correlated Matter, Zhejiang University, 
Hangzhou, Zhejiang 310058, China}\\
(Dated: December 8, 2016)\\[0.3cm]

\end{centering}

In this supplement we provide details of (1) the calculation of the spin 
scattering vertex between local $4f-$spins and conduction electrons, including
RKKY contributions and (2) the derivation of the one-loop RG equation and 
its solution.\\[0.8cm]  

\end{widetext}

\setcounter{equation}{0}
\setcounter{figure}{0}
\renewcommand{\theequation}{S\arabic{equation}}
\renewcommand{\thefigure}{S\arabic{figure}}


\section{f-spin -- conduction electron vertex $\hat\Gamma_{cf}$}
\label{sec:cfvertex}

The elementary $f-$spin -- $c-$electron vertex with coupling constant 
$J_0$ is defined via  the Kondo Hamiltonian, 
\be
H = \sum_{{\bf k},\sigma} 
\varepsilon_{\bf k} \; c_{{\bf k} \sigma}^\dagger c_{{\bf k} \sigma}^\phdagger +
J_0 \sum_{i} {\bf S}_i \cdot {\bf s}_i  \ ,
\label{S_Hamiltonian}
\ee
with the notation used in the main text. The direct ($d$) 
and exchange ($x$) parts of the RKKY-induced vertex can be written 
as the product of a distance and energy dependent 
function $\Lambda_{RKKY}^{(d/x)}$ and an operator in spin space, 
$\hat\gamma^{(d/x)}$,
\be 
\hat\gamma_{RKKY}^{(d/x)}= \Lambda_{RKKY}^{(d/x)}({\bf r}_{ij},i\Omega) \,
\hat\gamma^{(d/x)}
\label{eq:S_RKKYvertex}
\ee

\subsection{Spin structure}
\label{subsec:RKKYvertex-spin}

\begin{figure}[t]
\vspace*{0.3cm}
\includegraphics[width=0.85\linewidth]{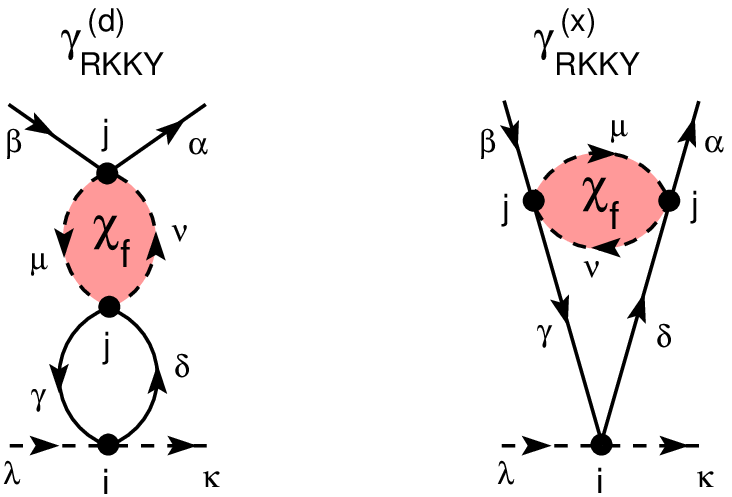}
\caption{\label{fig:S_RKKYvertex_spin} 
Direct ($d$) and exchange ($x$) diagrams 
of the RKKY-induced contributions to the $c-f$ vertex: spin labelling.}
\end{figure}

Denoting the vector of Pauli matrices acting in $c-$electron spin space 
by $\bsig =(\sigma^x,\sigma^y,\sigma^z)$ and the vector of Pauli matrices 
in $f-$spin space by ${\bf s}=(s^x,s^y,s^z)$,  
the RRKY-induced vertex contributions read in spin space,
\bea
\gamma^{(d)}_{\alpha\beta, \kappa\lambda} &=& \hspace*{-0.2cm}
\sum_{a,b,c=x,y,z}\ \sum_{\gamma,\delta,\mu,\nu=1}^2
\left( \sigma^{a}_{\delta\gamma}s^{a}_{\kappa\lambda}\right) 
\left( \sigma^{b}_{\gamma\delta}s^{b}_{\nu\mu}\right) 
\left( \sigma^{c}_{\alpha\beta}s^{c}_{\mu\nu}\right)\nonumber\\ 
\label{eq:S_gamma_spin-d}\\
\gamma^{(x)}_{\alpha\beta, \kappa\lambda} &=& \hspace*{-0.2cm}
\sum_{a,b,c=x,y,z}\ \sum_{\gamma,\delta,\mu,\nu=1}^2
\left( \sigma^{a}_{\delta\gamma}s^{a}_{\kappa\lambda}\right) 
\left( \sigma^{b}_{\alpha\delta}s^{b}_{\nu\mu}\right) 
\left( \sigma^{c}_{\gamma\beta}s^{c}_{\mu\nu}\right)\nonumber\\ 
\label{eq:S_gamma_spin-x}
\eea
with $c-$electron spin indices $\alpha$, $\beta$, $\gamma$, $\delta$, and 
$f-$spin indices $\kappa$, $\lambda$, $\mu$, $\nu$, 
as shown in Fig.~\ref{fig:S_RKKYvertex_spin}. The spin summations can be 
performed using the spin algebra ($a,b=x,y,z$), 
\bea
\sum_{\gamma=1}^2
\sigma^{a}_{\alpha\gamma}\sigma^{b}_{\gamma\beta} = \sum_{c=x,y,z}
i \varepsilon^{abc} \sigma^{c}_{\alpha\beta}
+\delta^{ab}\mathds{1}_{\alpha\beta} \ ,
\label{eq:S_spin_algebra}
\eea
where $\mathds{1}$ is the unit operator in spin space,   
$\varepsilon^{abc}$ the totally antisymmetric tensor and 
$\delta^{ab}$ the Kronecker-$\delta$. This results in a nonlocal 
Heisenberg coupling between sites $i$ and $j$,   
\bea
\gamma^{(d)}_{\alpha\beta, \kappa\lambda} &=& 
\phantom{+}4 \sum_{a=x,y,z}
\left( \sigma^{a}_{\alpha\beta}s^{a}_{\kappa\lambda}\right) 
\label{eq:S_gamma_spin_d2}\\
\gamma^{(x)}_{\alpha\beta, \kappa\lambda} &=& 
-2 \sum_{a=x,y,z}
\left( \sigma^{a}_{\alpha\beta}s^{a}_{\kappa\lambda}\right) 
\label{eq:S_gamma_spin_x2} \ .
\eea

\begin{figure}[t]
\includegraphics[width=0.85\linewidth]{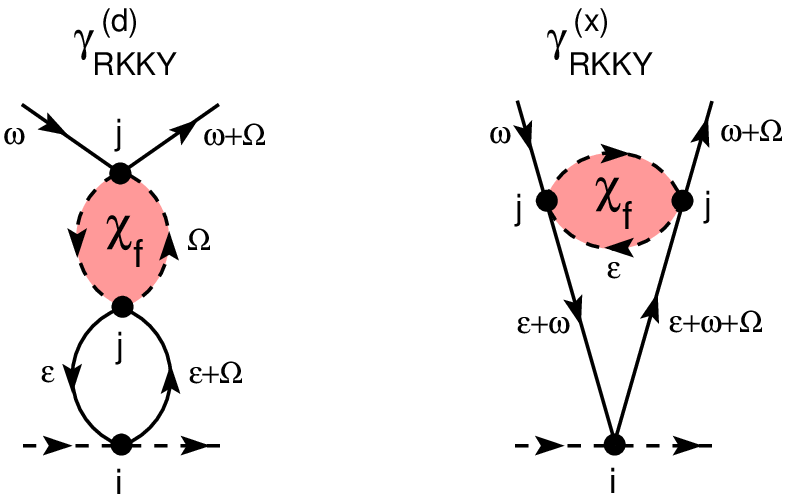}
\caption{\label{fig:S_RKKYvertex_freq} 
Direct ($d$) and exchange ($x$) diagrams 
of the RKKY-induced contributions to the $c-f$ vertex: energy labelling.}
\end{figure}

\subsection{Energy dependence}
\label{subsec:cfvertex-energy}

With the energy variables as defined in Fig.~\ref{fig:S_RKKYvertex_freq},
the energy dependent functions in \Eq{eq:S_RKKYvertex} read 
in Matsubara representation (the Matsubara indices are suppressed for 
simplicity of notation, i.e. $i\Omega \equiv i\Omega_n$, etc.),

\begin{widetext}
\bea
\Lambda^{(d)}_{RKKY}({\bf r}_{ij},i\Omega) &=& 
JJ_0^2\chi_c({\bf r}_{ij},i\Omega)\tilde\chi_f(i\Omega) 
\label{eq:S_Lambda_d0}\\ 
\Lambda^{(x)}_{RKKY}({\bf r}_{ij},i\omega,i\Omega) &=& 
-JJ_0^2 T\sum_{i\varepsilon} 
G_c({\bf r}_{ij},i\omega+i\varepsilon)
G_c({\bf r}_{ij},i\omega+i\Omega+i\varepsilon)\tilde\chi_f(i\varepsilon)
\label{eq:S_Lambda_x0}
\eea
where
\bea
\chi_c({\bf r}_{ij},i\Omega) &=& - T\sum_{\varepsilon}
G_c({\bf r}_{ij},i\varepsilon)G_c({\bf r}_{ij},i\varepsilon+i\Omega) \hspace*{2.1cm}
\label{eq:S_chic}
\eea
and $\tilde\chi_f(i\varepsilon)=\chi_f(i\varepsilon)/(g_L\mu_B)^2$, 
with $\chi_f(i\varepsilon)$ the full, single-impurity $f-$spin 
susceptibility whose temperature dependence is known from Bethe ansatz
(see main text).  
In an isotropic system in $d=3$ dimensions, the retarded
conduction electron Green's function $G_c$ as well as the 
susceptibilities $\chi_c$ and $\tilde\chi_f$ at temperature $T=0$ 
are calculated in position space as,
\bea
G_c({\bf r},\omega \pm i0) &=& - \pi N(\omega)\,
\frac{{\rm e}^{\pm i k(\varepsilon_F+\omega)r}}
{k(\varepsilon_F+\omega)r} 
\label{eq:S_Gc}\\
\chi_c({\bf r}_{ij},\Omega+i0) &=& 
\left[  N(0)\,
\frac{\sin(x)-x\cos(x)}{4x^4}\ + 
\ {\cal O}\left(\left(\frac{\Omega}{\varepsilon_F}\right)^2\right)
\right]
\pm i \left[\ \frac{1}{\pi} N(0)\, \frac{1-\cos(x)}{x^2}\, 
\frac{\Omega}{\varepsilon_F}\ 
+ \ {\cal O}\left(\left(\frac{\Omega}{\varepsilon_F}\right)^3\right)
\ \right]  \nonumber\\
\label{eq:S_chic_ret}\\
\tilde\chi_f(\Omega \pm i0) &=& 
\frac{W}{\pi D_0}\, \frac{D_0}{T_K}\,\frac{1}{\sqrt{1+(\Omega/T_K)^2}}\, 
\left( 1 \pm \frac{2i}{\pi} {\rm arsinh} \frac{\Omega}{T_K} \right)
\qquad\qquad \mathrm{with}\ \ A:=\frac{W}{\pi D_0} \ .
\label{eq:S_chif_ret}
\eea 
Here, $\varepsilon_F$ and $k_F$ are the Fermi energy and Fermi wavenumber, 
respectively, $x=2k_Fr$, $N(0)$ the conduction electron density of 
states at the Fermi energy, and $D_0$ the bare band cutoff.    

For the renormalization of the total $c-f$ vertex for $c-$electrons 
at the Fermi energy, the contributions $\Lambda^{(d)}_{RKKY}$, $\Lambda^{(d)}_{RKKY}$
must be calculated for real frequencies, $i\Omega\to \Omega+i0$,
$i\omega\to \omega+i0$, and for electrons at the Fermi energy, i.e.,
$\omega=0$. In this limit, only the real parts of 
$\Lambda^{(d)}_{RKKY}$, $\Lambda^{(d)}_{RKKY}$ contribute to the vertex 
renormalization, as shown below.
In order to analyze their importance for the RG flow, we will
expand them in terms of the small parameter $T_K/D_0$.
In the following, the real part of a complex function will be denoted 
by a prime {'} and the imaginary part by a double-prime {''}.\\[0.2cm]

{\bf Direct contribution} \\

Since in  $\Lambda^{(d)}_{RKKY}$ [\Eq{eq:S_Lambda_d0}], $\chi_c(i\Omega)$ and 
$\tilde\chi_f(i\Omega)$ appear as a product and $\chi_f(\Omega)$ 
cuts off the energy transfer $\Omega$ at the 
scale $T_K\ll \varepsilon_F\approx D_0$, the electron polarization 
$\chi_c(\Omega)$ contributes only in the limit $\Omega \ll \varepsilon_F$ 
where it is real-valued, as seen in \Eq{eq:S_chic_ret}. 
Using \Eq{eq:S_chic_ret} and \Eq{eq:S_chif_ret}, the real part of the 
direct RKKY-induced vertex contribution reads,
\bea
\Lambda^{(d)}_{RKKY}{'}({\bf r}_{ij},\Omega+i0) &=& 
JJ_0^2  R({\bf r}_{ij}) A N(0) \, 
\frac{D_0}{T_K}\,\frac{1}{\sqrt{1+(\Omega/T_K)^2}} \, +\,
{\cal O}\left(\left(\frac{\Omega}{D_0}\right)^2\right)\ ,
\label{eq:S_Lambda_d1}
\eea
where   
\bea   
R({\bf r}_{ij}) &=& \frac{\sin(x)-x\cos(x)}{4x^4}, \qquad\qquad x=2k_Fr
\eea   
is a spatially oscillating function.\\[0.3cm]
 
{\bf Exchange contribution}\\

In order to analyze the size of $\Lambda^{(x)}_{RKKY}{'}$ in terms of 
$T_K/D_0$, it is sufficient to evaluate it for a particle-hole symmetric 
conduction band and for ${\bf r}_{ij}=0$, since  
the $T_K/D_0$ dependence is induced by the on-site susceptibility
$\tilde\chi_f(i\Omega)$. The dependence on $T_K/D_0$ can be changed by 
the frequency convolution involved in $\Lambda^{(x)}_{RKKY}{'}$, but does not 
depend on details of the conduction band and distance dependent 
terms. (The general calculation is possible as well, but considerably 
more lengthy \cite{S_NejatiPhD}.)    
We use the short-hand notation for the momentum-integrated
$c-$electron Green's function,
$G_c({\bf r}=0,\omega \pm i0)=G(\omega)=G'(\omega)+iG''(\omega)$, and
assume a flat density of states $N(\omega)$, with the upper and lower
band cutoff symmetric about $\varepsilon_F$, i.e., 
\bea  
G^{R/A}{''}(\omega) &=& \mp \frac{\pi}{2D_0}\Theta(D_0-|\omega|)\\
G^{R/A}{'}(\omega)  &=& \frac{1}{2D_0} 
\ln\left|\frac{D_0+\omega}{D_0-\omega}\right| = \frac{\omega}{D_0^2} 
\,+\,{\cal O}\left( \left(\frac{\omega}{D_0}\right) \right) \ .
\eea   
Furthermore, at $T=0$ the Fermi and Bose distribution functions are,
$f(\varepsilon)=-b(\varepsilon)=\Theta(-\varepsilon)$.\\
$\Lambda^{(x)}_{RKKY}{'}(0,0,\Omega+i0)$ then reads, 
\bea
\Lambda^{(x)}_{RKKY}{'}({\bf r}_{ij}=0,\omega=0+i0,\Omega+i0) &=& \nonumber \\ 
&& \hspace*{-4cm}- JJ_0^2 \, \left\{  
\,\int \frac{d\varepsilon}{\pi}
\left[
f(\varepsilon) G^A{''}(\varepsilon)G^R{'}(\varepsilon+\Omega) 
+ f(\varepsilon+\Omega)\, G^A{'}(\varepsilon)G^A{''}(\varepsilon+\Omega)
\right] \, \tilde\chi_f^R{'}(\varepsilon) \right.
 \label{eq:S_Lambda_x}\\
&& \hspace*{-3.11cm}
\left. -\,\int \frac{d\varepsilon}{\pi}
\left[
f(\varepsilon)\, G^R{'}(\varepsilon)G^R{'}(\varepsilon+\Omega) 
- f(\varepsilon+\Omega) G^A{''}(\varepsilon)G^A{''}(\varepsilon+\Omega)
\right] \, \tilde\chi_f^R{''}(\varepsilon) \right\}
\nonumber \ .
\eea   
With the above definitions, the four terms in this expression are evaluated 
in an elementary way, using the substitution $\varepsilon_F/T_K=x=\sinh u$,
\bea
\int \frac{d\varepsilon}{\pi}
f(\varepsilon) G^A{''}(\varepsilon)G^R{'}(\varepsilon+\Omega) 
\tilde\chi_f^R{'}(\varepsilon) &=& 
A N(0) \frac{T_K}{D_0}
\left[
1-\sqrt{1+\left(\frac{D_0}{T_K}\right)^2}+
\frac{\Omega}{T_K}\, {\rm arsinh}\left(\frac{D_0}{T_K}\right)
\right] \nonumber\\
&=& 
A N(0) \left[
-1 + 
\frac{\Omega}{D_0}\,  \ln\left(\frac{D_0}{T_K}\right)
\, + \, 
{\cal O}\left(\frac{T_K}{D_0}\right)
\right]
\label{eq:S_Lambda_x1}
\eea  

\bea
\left|\int \frac{d\varepsilon}{\pi}
f(\varepsilon+\Omega)\, G^A{'}(\varepsilon)G^A{''}(\varepsilon+\Omega)
\tilde\chi_f^R{'}(\varepsilon) \right| &=&
A N(0) \frac{T_K}{D_0}
\left|
\sqrt{1+\left(\frac{\Omega}{T_K}\right)^2} -
\sqrt{1+\left(\frac{D_0+\Omega}{T_K}\right)^2} 
\right| \nonumber\\
&\leq& A N(0) \,+\, {\cal O}\left(\frac{T_K}{D_0}\right)
\label{eq:S_Lambda_x2}
\eea   

\bea
\int \frac{d\varepsilon}{\pi}
f(\varepsilon)\, G^R{'}(\varepsilon)G^R{'}(\varepsilon+\Omega) 
\tilde\chi_f^R{''}(\varepsilon) &=&
-\frac{4}{\pi^2} A N(0) \left(\frac{1}{2}+\frac{\Omega}{D_0} \right)\,
\ln\left(\frac{D_0}{T_K}\right) 
\,+\, {\cal O}\left(\left(\frac{T_K}{D_0}\right)^0\right)
\label{eq:S_Lambda_x3}
\eea   

\bea
\int \frac{d\varepsilon}{\pi}
f(\varepsilon+\Omega) G^A{''}(\varepsilon)G^A{''}(\varepsilon+\Omega)
\tilde\chi_f^R{''}(\varepsilon) &=& 
 \frac{\pi}{4}A N(0) \left[
- {\rm arsinh}\left(\frac{\Omega}{T_K}\right) 
+ {\rm arsinh}\left({\rm min}
\left(\frac{\Omega}{T_K},\frac{D_0+\Omega}{T_K}\right)\right) 
\right]\nonumber \\
&\leq&   \frac{\pi}{4}A N(0) 
\,+\, {\cal O}\left(\frac{T_K}{D_0}\right) \ .
\label{eq:S_Lambda_x4}
\eea   

Comparing  Eqs.~(\ref{eq:S_Lambda_x})--(\ref{eq:S_Lambda_x4}) with
\Eq{eq:S_Lambda_d1} shows that all terms of $\Lambda^{(x)}_{RKKY}{'}(\Omega)$ 
are subleading compared to $\Lambda^{(d)}_{RKKY}{'}(\Omega)$ by at least 
a factor $(T_K/D_0)\ln(T_K/D_0)$ for all transfered energies $\Omega$.
Hence, it can be neglected in the RG flow. 
Combining the results of spin and energy dependence, 
Eqs.~(\ref{eq:S_RKKYvertex}), (\ref{eq:S_gamma_spin_d2}) and (\ref{eq:S_Lambda_d1}), 
one obtains the total RKKY-induced $c-f$ vertex as, 
\bea 
\hat\gamma_{RKKY}^{(d)} ({\bf r}_{ij},i\Omega)&=&
2 \,(1-\delta_{ij})\,\chi_c({\bf r}_{ij},i\Omega)\chi_f(i\Omega)\,{\bf S}_i
\cdot {\bf s}_j
\eea
or
\bea 
{\rm Re}\hat\gamma_{RKKY}^{(d)} ({\bf r}_{ij},\Omega+i0)&=&
2 JJ_0^2 AN(0)\,(1-\delta_{ij}) R({\bf r}_{ij}) \,
\frac{D_0}{T_K}\,\frac{1}{\sqrt{1+(\Omega/T_K)^2}} 
\,{\bf S}_i \cdot {\bf s}_j \ .
\eea


\section{Perturbative renormalization group}
\label{sec:RG}

\subsection{One-loop RG equation}
\label{subsec:RGequation}

The derivation of the RG equation follows the well-known procedure 
of perturbative coupling constant renormalization \cite{S_Hewson93},
however performed for the nonlocal $c-f$ vertex including RKKY contributions.  
The amplitude of the  
sum of the diagrams contributing to the one-loop renormalization of 
the $c-f$ vertex reads in Matsubara representation 
(c.f. Fig.~1 (b) and Eq.~(4) of the main paper),
\begin{eqnarray}
Y({\bf r}_{ij}, i\omega ) =   \label{eq:S_Yx} 
&-&T \sum_{i\Omega}   
\left[  J^2\delta_{ij} +
J\gamma^{(d)}_{RKKY}({\bf{r}}_{ij},i\Omega)+
J\gamma^{(d)}_{RKKY}({\bf{r}}_{ij},-i\Omega)
\right] G_c({\bf r}_{ij}, i\omega -i\Omega)\,G_f(i\Omega) \\
&+&T \sum_{i\Omega}   
\left[  J^2\delta_{ij} +
J\gamma^{(d)}_{RKKY}({\bf{r}}_{ij},i\Omega)+
J\gamma^{(d)}_{RKKY}({\bf{r}}_{ij},-i\Omega)
\right] G_c({\bf r}_{ij}, i\omega +i\Omega)\,G_f(i\Omega) 
\nonumber
\end{eqnarray}
This is a nonlocal function of the ingoing and outgoing coordinates 
of $c-$electrons, $x_j$, $x_i$. For the low-energy strong coupling 
fixed point the coupling constant for $c-$electrons at the Fermi energy
must be renormalized, i.e., for excitation energy 
$\omega=0$ and momentum on the Fermi surface, ${\bf k}_F$. 
Hence, the coupling constant renormalization is given by the Fourier 
transform of $Y({\bf r}_{ij}, \omega=0+i0)$ with respect to the 
in- and outgoing coordinates, ${\bf x}_j$, ${\bf x}_i$, taken for 
momenta ${\bf k}_{in}$, ${\bf k}_{out}$ on the Fermi surface. 
In a lattice system, translation invariance implies the 
conservation of in in- and outgoing momenta, 
${\bf k}_{in}={\bf k}_{out}={\bf k}_F$. This yields,
\begin{eqnarray}
Y({\bf k}_{F}, i\omega ) =   \label{eq:S_Yk1} 
&-&J^2\, T \sum_{i\Omega}   
\left[
 G_c({\bf r}_{ij}\to 0, i\omega -i\Omega)\,G_f(i\Omega)
-G_c({\bf r}_{ij}\to 0, i\omega +i\Omega)\,G_f(i\Omega)
\right]\\
&-&J\, T \sum_{i\Omega}
\sum_j {\rm e}^{+i{\bf k}_F{\bf r}_{ij}}\,
\left[
\gamma^{(d)}_{RKKY}({\bf{r}}_{ij},i\Omega)+
\gamma^{(d)}_{RKKY}({\bf{r}}_{ij},-i\Omega)
\right] G_c({\bf r}_{ij}, i\omega -i\Omega )\,G_f(i\Omega)\nonumber \\
&+&J\,T \sum_{i\Omega}   
\sum_j {\rm e}^{-i{\bf k}_F{\bf r}_{ij}}\,
\left[ 
\gamma^{(d)}_{RKKY}({\bf{r}}_{ij},i\Omega)+
\gamma^{(d)}_{RKKY}({\bf{r}}_{ij},-i\Omega)
\right] G_c({\bf r}_{ij}, i\omega +i\Omega )\,G_f(i\Omega)\ ,\nonumber
\end{eqnarray}
where the $j-$summation runs over all lattice sites $j$ which are occupied
by a Kondo ion.  
Using the symmetry 
$
\gamma^{(d)}_{RKKY}({\bf{r}}_{ij},-\Omega)= 
\gamma^{(d)*}_{RKKY}({\bf{r}}_{ij},\Omega)
$, and the pseudofermion propagator $G_f(i\Omega)=1/i\Omega$,
projected onto the physical Hilbert space \cite{S_Kroha98},
the frequency summations can be performed to yield,
\begin{eqnarray}
Y({\bf k}_{F}, \omega =0) =   \label{eq:S_Yk2} 
&-& 2 J^2 N(0)
\left[
\int_{T}^{D}  \frac{d\Omega}{\Omega} \, - \,
\int_{-D}^{-T} \frac{d\Omega}{\Omega} 
\right]
\\ 
&-& 2 J \frac{1}{\pi} \int_T^D\,\,\, \frac{d\Omega}{\Omega} \,
{\rm Im} \left[
\sum_{j} {\rm e}^{+i{\bf k_F}{\bf r}_{ij}}\,
2\gamma^{(d)}_{RKKY}{'}({\bf{r}}_{ij},\Omega) \, 
G_c({\bf r}_{ij}, \Omega -i0)  
\right] \nonumber \,
\end{eqnarray}
Where $D$ is the running band cutoff. The change of 
$Y({\bf k}_{F}, \omega =0)$ under an infinitesimal, logarithmic 
cutoff reduction $d\ln D$ represents the renormalization of the 
$c-f$ coupling constant. That is, the renormalization group 
equation is obtained as, 
\begin{eqnarray}
\frac{dg}{d\ln D} =
\frac{dY({\bf k}_{F}, \omega =0)N(0)}{d\ln D} =   \label{eq:S_RG1} 
\,-\, 2g^2 
\,-\, \frac{4}{\pi} g\, 
{\rm Im} \left[
\sum_{j} {\rm e}^{-i{\bf k_F}{\bf r}_{ij}}\,
\gamma^{(d)}_{RKKY}{'}({\bf{r}}_{ij},D) \, 
G_c({\bf r}_{ij}, D +i0)  
\right] \ ,
\end{eqnarray}
\end{widetext}
where $g=JN(0)$, $g_0=J_0N(0)$ are the dimensionless couplings.
Defining the RKKY coupling parameter as,
\be
y=-\frac{8W}{\pi ^2} {\rm Im} \sum_{j\neq i}
\frac{{\rm e}^{-i{\bf k}_F{\bf r}_{ij}}} {N(0)^2}
 G_c^R({\bf r}_{ij}, \Omega=0 ) \,
\chi_c({\bf r}_{ij}, \Omega=0 ) \ ,
\label{eq:S_RKKYparameter1}
\ee
\begin{figure}[t]
\includegraphics[width=0.85\linewidth]{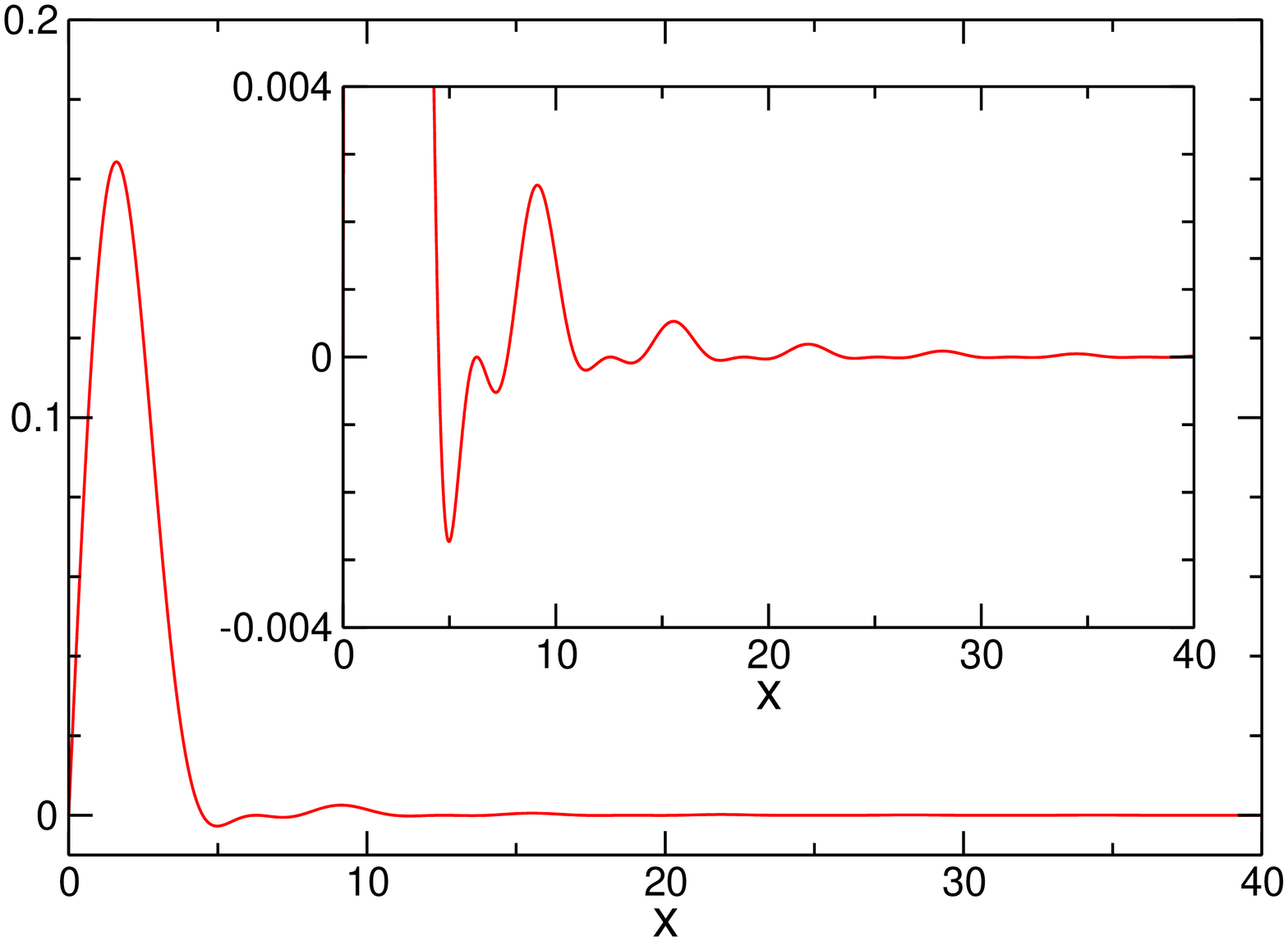}
\caption{\label{fig:S_y_integ} 
Integrand of the expression for $y$, \Eq{eq:S_RKKYparameter2}.
It illustrates that $y>0$ and that $y$ depends sensitively on $k_Fa$.
The inset shows the curve on a smaller scale.}
\end{figure}
\hspace*{-0.1cm}the RG equation takes the form,
\begin{eqnarray}
\frac{dg}{d\ln D}  =   \label{eq:S_RG2} 
\,-\, 2g^2 \left[
1-y g_0^2\, \frac{D_0}{T_K}\,\frac{1}{\sqrt{1+(\Omega/T_K)^2}} 
\right] \ .
\eea
It naturally reduces to the single-impurity Kondo RG equation, if the 
RKKY interaction is switched off ($y=0$).
In a dense Kondo lattice with lattice constant $a$ in $d=3$ dimensions, 
$k_Fa\ll 1$, the lattice summation in \Eq{eq:S_RKKYparameter1} can be 
approximated by an integral, and $y$ becomes,
\be
y \approx  \frac{2W}{(k_Fa)^3}
\int_{k_Fa}^{\infty} dx\ (1-\cos x)\ \frac{x \cos x - \sin x}{x^4} > 0 \ .
\label{eq:S_RKKYparameter2}
\ee 
$y$ depends sensitively on $k_Fa$. It represents the
dependence on the lattice structure of the Kondo ions ($a$) and 
on the band filling ($k_f$). For illustration we show in 
Fig.~\ref{fig:S_y_integ} the integrand of the expression 
for $y$, \Eq{eq:S_RKKYparameter2}.
It is also seen that $y>0$, i.e. the RKKY coupling always {\it reduces} the 
effective coupling strength between conduction electrons and 
local $f-$spins, irrespective of the oscillating sign of the 
RKKY correlations, 
\Eq{eq:S_chic_ret}. This is physically expected, since a  
ferro- as well as an antiferromagnetic coupling of an $f-$spin to the
neighboring $f-$spins will always reduce the local spin fluctuations 
and, therefore, the Kondo singlet formation. For a non-translational
invariant systems, like two- or multi-impurity Kondo systems, the 
expression for $y$ will somewhat differ from 
\Eq{eq:S_RKKYparameter1}, since in- and outgoing momenta are not conserved. 
However, this does not change the form of the RG equation 
(\ref{eq:S_RG2}).\\[0.3cm]

\subsection{Integration of the RG equation}
\label{subsec:RGintegration}

The RG equation \Eq{eq:S_RG2} is readily integrated by separation of variables,
\bea
-\int_{g_0}^{g}\frac{dg}{g^2} &=&
2 \int_{\ln D_0}^{\ln D} d\ln D' 
\label{eq:S_RGint1} \\
&-&2yg_0^2\frac{D_0}{T_K}\int_{D_0/T_K}^{D/T_K}\frac{dx}{x}
\frac{1}{\sqrt{1+x^2}} \ ,
\nonumber
\eea 
or 
\bea
\frac{1}{g}-\frac{1}{g_0} &=& \label{eq:S_RGint2}
2 \ln\left(\frac{D}{D_0}\right) \\
&-&yg_0^2\frac{D_0}{T_K} 
\ln\left(
\frac{\sqrt{1+(D/T_K)^2}-1}{\sqrt{1+(D/T_K)^2}+1}
\right)
\nonumber 
\eea 
where we have used $D_0/T_K\gg 1$ in the last expression.

The Kondo scale is defined as the value of the running cutoff $D$
where $g$ diverges, i.e., $g\to\infty$  when $D\to T_K$. This 
yields the defining equation for the Kondo scale which, hence, depends 
on the RKKY parameter, $T_K\equiv T_K(y)$,
\bea
-\frac{1}{g_0} &=& \label{eq:S_TKydef1}
2 \ln\left(\frac{T_K(y)}{D_0}\right) \\
&-&yg_0^2\frac{D_0}{T_K(y)} 
\ln\left(
\frac{\sqrt{2}-1}{\sqrt{2}+1}
\right) 
\nonumber
\eea 
Using the definition of the single-impurity Kondo temperature,
$-1/g_0=2\ln\left(T_K(0)/D_0\right)$, the defining equation 
for $T_K(y)$ can finally be written as 
\bea
\frac{T_K(y)}{T_K(0)}=
{\rm exp}\left(-y\alpha g_0^2\frac{D_0}{T_K(y)}\right) \ ,
\eea 
with $\alpha=\ln(\sqrt{2}+1)$.



\begin{thebibliography} {99}

\bibitem{Loehneysen07}
H. v. L\"ohneysen, A. Rosch, M. Vojta and P. W\"olfle, 
Fermi-liquid instabilities at magnetic quantum phase transitions,
Rev. Mod. Phys. {\bf 79}, 1015 (2007).

\bibitem{Hewson93} 
A. C. Hewson, {\it The Kondo Problem to Heavy Fermions}
(Cambridge University Press, Cambridge, England, 1993).

\bibitem{Hertz76}
J. A. Hertz, 
Quantum critical phenomena,
Phys. Rev. B {\bf 14}, 1165 (1976).

\bibitem{Millis93}
A. Millis, 
Effect of a nonzero temperature on quantum critical points in itinerant 
fermion systems,
Phys. Rev. B {\bf 48}, 7183 (1993).

\bibitem{Moriya85}
T. Moriya, {\it Spin Fluctuations in Itinerant
Electron Magnetism} (Springer, Berlin, 1985).

\bibitem{Doniach77}
S. Doniach, 
Kondo lattice and weak antiferromagnetism,
Physica B+C {\bf 91}, 231 (1977).

\bibitem{Ruderman54}
M. A. Ruderman and C. Kittel, 
Indirect exchange coupling of nuclear magnetic moments by conduction electrons,
Phys. Rev. {\bf 96}, 99 (1954).

\bibitem{Kasuya56}
T. Kasuya, 
A Theory of Metallic Ferro- and Antiferromagnetism on Zener's Model,
Prog. Theor. Phys. {\bf 16}, 45 (1956);

\bibitem{Yosida57}
K. Yosida, 
Magnetic properties of Cu-Mn alloys,
Phys. Rev. {\bf 106}, 893 (1957).


\bibitem{Si01}
Q. Si, S. Rabello, K. Ingersent  and J. L. Smith,  
Locally critical quantum phase transitions in strongly correlated metals,
Nature (London) {\bf 413}, 804 (2001).

\bibitem{Coleman01}
P. Coleman, C. P\'epin, Q. Si and R. Ramazashvili, 
How do Fermi liquids get heavy and die?, 
J. Phys. Condens. Matter {\bf 13}, R723 (2001).


\bibitem{Senthil04}
T. Senthil, M. Vojta and S. Sachdev, 
Weak magnetism and non-Fermi liquids near heavy-fermion critical points,
Phys. Rev. B {\bf 69}, 035111 (2004).


\bibitem{Woelfle11}
P. W\"olfle and E. Abrahams, 
Quasiparticles beyond the Fermi liquid and heavy fermion criticality,
Phys. Rev. B {\bf 84}, 041101(R) (2011).

\bibitem{Woelfle14}
E. Abrahams, J. Schmalian, and P. W\"olfle,  
Strong-coupling theory of heavy-fermion criticality,
Phys. Rev. B {\bf 90}, 045105 (2014).

\bibitem{Woelfle16}
P. W\"olfle and E. Abrahams,  
Vertex functions at finite momentum: 
Application to antiferromagnetic quantum criticality,
Phys. Rev. B {\bf 93}, 075128 (2016).

\bibitem{Affleck95}
I. Affleck, A. W. W. Ludwig, and B. A. Jones,
Conformal field theory approach to the two-impurity Kondo problem:
comparison with numerical renormalization group results,
Phys. Rev. B {\bf 52}, 9528 (1995).


\bibitem{Sela09}
E. Sela and I. Affleck, 
Resonant Pair Tunneling in Double Quantum Dots,
Phys. Rev. Lett. {\bf 103}, 087204 (2009).


\bibitem{Bork11}
J. Bork, Y.-H. Zhang, L. Diekh\"oner, L. Borda, P. Simon, J. Kroha, 
P. Wahl, and K. Kern, 
A tunable two-impurity Kondo system in an atomic point contact,
Nature Physics {\bf 7}, 901 (2011).

\bibitem{Kroeger11}
N. N\'eel, R. Berndt, J. Kr\"oger, T. O. Wehling, A. I. Lichtenstein,
and M. I. Katsnelson,
Two-Site Kondo Effect in Atomic Chains,
Phys. Rev. Lett. {\bf 107}, 106804 (2011).


\bibitem{Prueser14}
H. Pr\"user, P. E. Dargel, M. Bouhassoune, R. G. Ulbrich, T. Pruschke,
S. Lounis, and M. Wenderoth,
Interplay between the Kondo effect and the 
Ruderman-Kittel-Kasuya-Yosida interaction,
Nat. Commun. {\bf 5}, 5417 (2014).

\bibitem{Friedemann09}
S. Friedemann,  T. Westerkamp, M. Brando, N. Oeschler, S. Wirth, P. Gegenwart, 
C. Krellner, C. Geibel, and F. Steglich, 
Detaching the antiferromagnetic quantum critical point from the 
Fermi-surface reconstruction in YbRh$_2$Si$_2$,
Nat. Phys. {\bf 5}, 465 (2009).


\bibitem{Klein08}
M. Klein, A. Nuber, F. Reinert, J. Kroha, O. Stockert and H. v. L\"ohneysen,
Signature of Quantum Criticality in Photoemission Spectroscopy, 
Phys. Rev. Lett. {\bf 101}, 266404 (2008).


\bibitem{Jones88}
B. A. Jones, C. M. Varma, and J. W. Wilkins, 
Low-Temperature Properties of the Two-Impurity Kondo Hamiltonian,
Phys. Rev. Lett. {\bf 61}, 125 (1988).



\bibitem{supplement}
See Supplemental Material 
for details of the calculations, which includes Ref.~\cite{Nejati16a}.



\bibitem{Nejati16a}
A. Nejati, PhD thesis, University of Bonn, Germany, 2016. 




\bibitem{Kroha98}
J. Kroha and P. W\"olfle,
Fermi and non-Fermi liquid behavior in quantum impurity systems: 
conserving slave boson theory, 
Acta Phys. Polonica B {\bf 29}, 3781 (1998).


\bibitem{Andrei83}
N. Andrei, K. Furuya, and J. H. Lowenstein, 
Solution of the Kondo problem,
Rev. Mod. Phys. {\bf 55}, 331 (1983).



\bibitem{Lambert} 
D. Veberi\'c, 
Lambert W function for applications in physics,
Comput. Phys. Commun. {\bf 183}, 2622 (2012); arXiv:1209.0735.

\bibitem{Nejati16}
A. Nejati and J. Kroha, 
Oscillation and suppression of Kondo temperature by RKKY coupling 
in two-site Kondo systems, arXiv:1612.06620
[J. Phys.: Conf. Ser. (to be published)].


\end{thebibliography}

\begin{thebibliography} {99}

\bibitem{S_Hewson93} 
A. C. Hewson, {\it The Kondo Problem to Heavy Fermions,
Cambridge University Press}, Cambridge, UK (1993).
 
\bibitem{S_NejatiPhD} 
A. Nejati, {\it Quantum phase transitions in multi-impurity and lattice 
Kondo systems}, PhD thesis, University of Bonn, Germany (2016).
 
\bibitem{S_Kroha98}
J. Kroha and P. W\"olfle,
Fermi and non-Fermi liquid behavior in quantum impurity systems: 
conserving slave boson theory, 
Acta Phys. Polonica B {\bf 29}, 3781 (1998);
arXiv:cond-mat/9811074.
\end{thebibliography}
\end{document}